\documentclass[a4paper,10pt,notitlepage]{revtex4-1}
\usepackage{amsmath,amssymb,textcomp,float,dsfont}
\usepackage[utf8]{inputenc}
\usepackage{amsmath}
\usepackage{amsfonts}
\usepackage{amssymb}
\usepackage{times,fullpage}
\usepackage{comment,graphicx}
\usepackage{array}

\begin{document}
\title{Transient exchange fluctuation theorems for heat using Hamiltonian framework: Classical and Quantum}
\author{P. S. Pal$^{a,b}$\email{}, Sourabh Lahiri$^c$\email{} and A. M. Jayannavar$^{a,b}$\email{}} 
\email{priyo@iopb.res.in, sourabhlahiri@gmail.com, jayan@iopb.res.in}
\affiliation{$^a$Institute of Physics, Sachivalaya Marg, Bhubaneswar-751005, India\\$^b$Homi Bhabha National Institute, Training School Complex, Anushakti Nagar, Mumbai 400085, India\\$^c$ International Centre for Theoretical Sciences (ICTS), Survey No. 151, Shivakote, Hesaraghatta Hobli, Bengaluru - 560 089, India.}

\begin{abstract}
We investigate the statistics of heat exchange between a finite system coupled to reservoir(s). We have obtained analytical results for  heat fluctuation theorem in the transient regime considering the Hamiltonian dynamics of the composite system consisting of the system of interest and the heat bath(s). The system of interest is driven by an external protocol. We first derive it in the context of a single heat bath. The result is in exact agreement with known result.  We then generalize the treatment to two heat baths. We further extend the study to quantum systems and show that relations similar to the classical case hold in the quantum regime. For our study we invoke von Neumann two point projective measurement  in quantum mechanics in the transient regime. Our result is a generalisation of Jarzynski-W$\grave{o}$jcik heat fluctuation theorem.
\end{abstract}
\maketitle{}
\newcommand{\nwc}{\newcommand}
\nwc{\vs}{\vspace}
\nwc{\hs}{\hspace}
\nwc{\la}{\langle}
\nwc{\ra}{\rangle}
\nwc{\nn}{\nonumber}
\nwc{\Ra}{\Rightarrow}
\nwc{\wt}{\widetilde}
\nwc{\lw}{\linewidth}
\nwc{\ft}{\frametitle}
\nwc{\ben}{\begin{enumerate}}
\nwc{\een}{\end{enumerate}}
\nwc{\bit}{\begin{itemize}}
\nwc{\eit}{\end{itemize}}
\nwc{\dg}{\dagger}
\nwc{\mA}{\mathcal A}
\nwc{\mD}{\mathcal D}
\nwc{\mB}{\mathcal B}
\nwc{\Tr}[1]{\underset{#1}{\mbox{Tr}}~}
\nwc{\pd}[2]{\frac{\partial #1}{\partial #2}}
\nwc{\ppd}[2]{\frac{\partial^2 #1}{\partial #2^2}}
\nwc{\fd}[2]{\frac{\delta #1}{\delta #2}}
\nwc{\pr}[2]{K(i_{#1},\alpha_{#1}|i_{#2},\alpha_{#2})}
\nwc{\av}[1]{\left< #1\right>}
\section{Introduction}
The fluctuation theorems (FTs) \cite{rit05_pt,jar11_ann_rev,jar97_prl,jar97_pre, harris07_jstat, rit03,han11_rmp,rana12_pramana,rana13_pramana,sei12,lah15_ijp} are a group of exact relations that remain valid even when the system of interest is driven far away from equilibrium. For driven systems fluctuations in heat, work and entropy are not mere background noise, but satisfy strong  constraints on the probability distributions of these fluctuating quantities.These relations are of fundamental importance in non-equilibrium statistical mechanics. Intensive research has been done in this direction in order to find such relations for thermodynamic quantities like work, heat or entropy changes. They have resulted in conceptual understanding of how irreversibility emerges from reversible dynamics and the second law of Thermodynamics. The second law holds for average quantities. However, there are atypical transient trajectories which violate second law \cite{mam11_jpa}. Two fundamental ingredients play a decisive role in the foundations of FT - the principle of microreversibility and the fact that thermal equilibrium is described by Gibbs canonical ensemble. Moreover, some of these relations have been found useful for practical applications like determining the change in equilibrium free energy in an irreversible process \cite{jar97_prl,jar97_pre}. Numerous FTs have been put forward in the last two decades. Some of them are valid when the system is in a nonequilibrium steady state \cite{lah14_epjb}, while the others are valid in the transient regime. Fluctuation theorems have been proposed for Hamiltonian \cite{jar11_ann_rev, jar07_crp,deffner13_prx} as well as stochastic dynamics\cite{rit05_pt,jar11_ann_rev,jar97_prl,jar97_pre, harris07_jstat, rit03,han11_rmp,rana12_pramana,rana13_pramana,sei12,lah15_ijp}, and for quantum systems \cite{rana12_pramana,rana13_pramana}(both closed and open ones) (for comprehensive review see \cite{han11_rmp} and references therein). Some of them have been tested experimentally \cite{rit05_pt,rit04_pnas, colin05_nat, sch05_prl, shu15_nat}.

One of these FTs has been that of the heat exchanged between two systems at different temperatures placed in direct contact with each other \cite{jar04_prl}. There it was shown that the heat exchanged between the two systems follows an exchange fluctuation theorem (XFT). The extension of this theorem to the non-ideal case where the contact between the two systems is through a heat conductor has been studied in \cite{lah14_epjb,gom06_pre}. 

The XFT in the presence of heat conductor has been studied for systems that are governed by the Langevin equation of motion \cite{lah14_epjb}. We show in this article that the XFT can also be derived in a Hamiltonian framework, where we consider the composite system containing the system as well as the two heat baths (maintained at different temperatures) to be governed by Hamiltonian dynamics. However, we need to assume that the interaction Hamiltonian between the system and the heat baths is small (weak coupling limit) compared to the Hamiltonians of the system or of the heat baths. This assumption helps us in having clear definitions of thermodynamic quantities like dissipated heat, work done or the change in internal energy. Nevertheless, the interaction terms are needed to ensure that the combined system undergoes deterministic evolution.

We further extend the treatment to quantum systems, where work, heat and internal energy  are defined through von Neumann two point projective measurements on the combined system, one at the beginning and the other performed at the end of the protocol. We show that relations similar to those of the classical system hold in the quantum regime as well. Our result generalizes the earlier  heat FT by Jarzynski and W$\grave{o}$jcik (J-W) in presence of an external protocol.

In section \ref{sec:single}, we describe the formalism for the simple case of a system in contact with a single heat bath. In section \ref{sec:system}, we describe our system and provide definitions of the thermodynamic quantities. In section \ref{sec:central}, we provide the derivation of our central result. In section \ref{sec:quantum}, we extend these derivations to the case of quantum systems. Finally, we conclude in section \ref{sec:conclusion}.

\section{Classical Treatment}

\subsection{System connected single heat bath}
\label{sec:single}
We consider a system connected to a single heat bath of temperature $T$. The full system is described by Hamiltonian dynamics. In the following, the term \emph{system} without being preceded by any adjective will imply the system of interest and not the composite system. Let a point in the system's phase space be denoted by $z$ and that of the heat bath be denoted by $y$. The system is subjected to external protocol $\lambda(t)$.The total Hamiltonian is given by
 \begin{align}
 H_{tot}(z,y;\lambda(t))&= H_s(z,\lambda(t))+H_{b}(y)+H_{int}(z,y).
 \end{align}
Here $H_s$ and $H_b$ are the Hamiltonians of the system and the heat bath respectively. $H_{int}$ denotes the interaction between the bath and system. This interaction term is assumed to be negligible compared to the other terms of the Hamiltonian\cite{deffner13_prx, han11_rmp}. However, the term must be present in order to ensure the Hamiltonian evolution of the full system. The heat bath is initially prepared at temperatures $T$ i.e., initial microstate $y(0)$ is sampled from canonical ensemble:
\begin{align}
\label{dist}
 P_b(y(0)) &= \frac{e^{-\beta H_{b}(y(0))}}{Z_b}.
 \end{align}
 Here, $Z_b$ is the canonical partition function for the heat bath.
The system is initially prepared in a state of equilibrium with this bath, and its probability distribution at initial time is given by
\begin{align}
  p_s(z(0)) &= \frac{e^{-\beta H_s(z(0))}}{Z_s(0)}.
\end{align}
At time $t=0$, we assume that the initial distribution of the combined system is given by
\begin{align}
P(\Gamma(0)) &= P_b(y(0))p_s(z(0)).
\end{align}
 At time $t=0^+$,  the external protocol is switched on.

 During the process, the composite system undergoes an Hamiltonian evolution (forward process denoted by $\Pi^+$) from the initial state $\Gamma=(z(0),y(0))$ and let 
\begin{align}
\hat\Gamma^t_+(\Gamma)\equiv(\hat z^t_+(\Gamma), \hat y^t_{+}(\Gamma))
\end{align}
denote the point in phase space reached after time $t$. The system is equilibrated with the bath at the end of the process. Let the final time (after equilibration) be $\tau$. At $t=\tau$, the final microstate of the composite system in the full phase space is $\Gamma'=\hat\Gamma^{\tau}_+(\Gamma)\equiv(\hat z^{\tau}_+(\Gamma), \hat y^{\tau}_{+}(\Gamma))=(z',y')$.

The thermodynamic quantities are defined as follows. The work done on the system equals the change in the total Hamiltonian of the full system (system+reservoir) and is given by
\begin{align}
  W = H_{tot}(\tau)-H_{tot}(0) = \int_0^\tau  \partial_tH_s(t) dt.
\end{align}
The arguments of the Hamiltonians have been suppressed for notational simplicity.
The heat dissipated into the reservoir must be the change in the Hamiltonian of the reservoir:
\begin{align}
  Q = H_b(\tau)-H_b(0).
\end{align}
The change in internal energy of the system is 
\begin{align}
  \Delta E = H_s(\tau)-H_s(0).
\end{align}

We now calculate the joint probability $P_+(Q,\Delta E)$ \cite{garcia10_pre} in the forward process
 \begin{eqnarray}
 P_+(Q,\Delta E)&=&\langle\delta(Q-\hat Q^+(\Gamma))\delta(\Delta E-\Delta \hat E^+(\Gamma))\rangle\nonumber\\
 &=&\int dz(0)dy(0) p_s(z(0))P_b(y(0))\delta(Q-\hat Q^+(\Gamma))\delta(\Delta E-\Delta \hat E^+(\Gamma)).
 \label{joint_single}
 \end{eqnarray}
 
Now we can rewrite $p_s(z(0))$ and $P_b(y(0))$ as 
\begin{eqnarray}
p_s(z(0))&=&e^{\beta[H_s(z')-H_s(z(0))]}.\frac{Z_s(\tau)}{Z_s(0)}.\frac{e^{-\beta H_s(z')}}{Z_s(\tau)}=\frac{e^{-\beta H_s(z')}}{Z_s(\tau)}e^{\beta(\Delta E-\Delta F)},\\
P_b(y(0))&=&\frac{P_b(y(0))}{P_b(y')}P_b(y')=e^{\beta Q}P_b(y').
\end{eqnarray}
where $\beta$ is the inverse of temperature $T$ ($\beta=1/k_BT$, we have set Boltzmann constant $k_B$ to unity) of the heat bath, and $\Delta F = \ln\frac{Z_s(\tau)}{Z_s(0)}$ is the change in the system's equilibrium free energy.  Substituting these expressions in Eq.\eqref{joint_single} and using the fact that the Jacobian between $(z(0),y(0))$ and $(z',y')$ is unity (by Liouville's theorem), we have

\begin{equation}
 P_+(Q,\Delta E)=\int e^{\beta [Q+\Delta E-\Delta F]} dz'dy' \frac{e^{-\beta H_s(z')}}{Z_s(\tau)}P_b(y')\delta(Q-\hat Q^+(\Gamma))\delta(\Delta E-\Delta \hat E^+(\Gamma)).
 \label{joints_c}
\end{equation}
Now if we reverse the final momenta and start a realisation of $\Pi^-$ from the initial conditions $\Gamma'^*$ , we will obtain the time-reversed image of the original realisation: $\hat\Gamma^t_-(\Gamma'^*)=[\hat\Gamma^{\tau-t}_+(\Gamma)]^*$. It should be clear that in the reverse process initially the bath microstate are chosen from the canonical ensemble given by
\begin{align}
 P_b(y'^*) &= \frac{e^{-\beta H_{b}(y'^*)}}{Z_b}=\frac{e^{-\beta H_{b}(y')}}{Z_b}=P_b(y'),
 \label{ris_c1}
 \end{align}
 and the initial system microstate is chosen from 
 \begin{align}
   p_s(z'^*)&= \frac{e^{-\beta H_s(z'^*)}}{Z_s(\tau)}=\frac{e^{-\beta H_s(z')}}{Z_s(\tau)}.
   \label{ris_c2}
 \end{align}
  From time reversal it follows that 
 \begin{align}
\hat Q_h^-(\Gamma'^*)=-\hat Q_h^+(\Gamma); \hspace{.3cm}\Delta \hat E^-(\Gamma'^*)=-\Delta \hat E^+(\Gamma).
\label{qs_c}
 \end{align}
 Eqs. \ref{ris_c1}, \ref{ris_c2} and \ref{qs_c} allows us to rewrite Eq.\ref{joints_c} as 
 
 \begin{eqnarray}
 P_+(Q,\Delta E)&=&e^{\beta [Q+\Delta E-\Delta F]}\int dz'^*dy'^* p_s(z'^*)P_b(y'^*)\delta(Q_h+\hat Q_h^-(\Gamma'^*)) \delta(\Delta E+\Delta \hat E^-(\Gamma'^*))\nonumber\\
 &=&e^{\beta [Q+\Delta E-\Delta F]}P_-(-Q,-\Delta E).
 \end{eqnarray}
Re-arranging the above equation and integrating over all possible values of $\Delta E$ we get
 \begin{eqnarray}
 \int e^{-\beta [Q+\Delta E-\Delta F]}P_+(Q,\Delta E)d(\Delta E)&=&\int P_-(-Q,-\Delta E)d(\Delta E),\nonumber\\
 e^{-\beta Q} \int e^{-\beta [\Delta E-\Delta F]}P_+(\Delta E|Q)d(\Delta E)P_+(Q)&=&P_-(-Q),\nonumber\\
\frac{P_+(Q)}{P_-(-Q)}=\frac{e^{\beta Q}}{\Psi(Q)}, 
\label{hfs_c}
 \end{eqnarray}

 where
 \begin{equation}
 \Psi(Q)=e^{\beta\Delta F}\int e^{-\beta\Delta E} P_+(\Delta E|Q)d(\Delta E). 
 \label{psi_classical}
 \end{equation}
 This is in agreement with \cite{noh12_prl} apart from an extra factor of free energy change ($\Delta F$) which arises due to the definition of inclusive thermodynamic work in our study.
 
%\subsection{Quantum} 

\subsection{System connected to two heat baths}
\label{sec:system}

%\subsection{The system}
We consider a composite system that is composed of a system connected to two different heat baths at different temperatures $T_h$ and $T_c$.  Let a point in the system's phase space be denoted by $z$, and those of the hot and the cold baths be denoted by $y_h$ and $y_c$, respectively. We will follow the treatment given in \cite{jar00_jsp} in our analysis.
\begin{figure}[H]
\begin{center}
 \includegraphics[width=8cm]{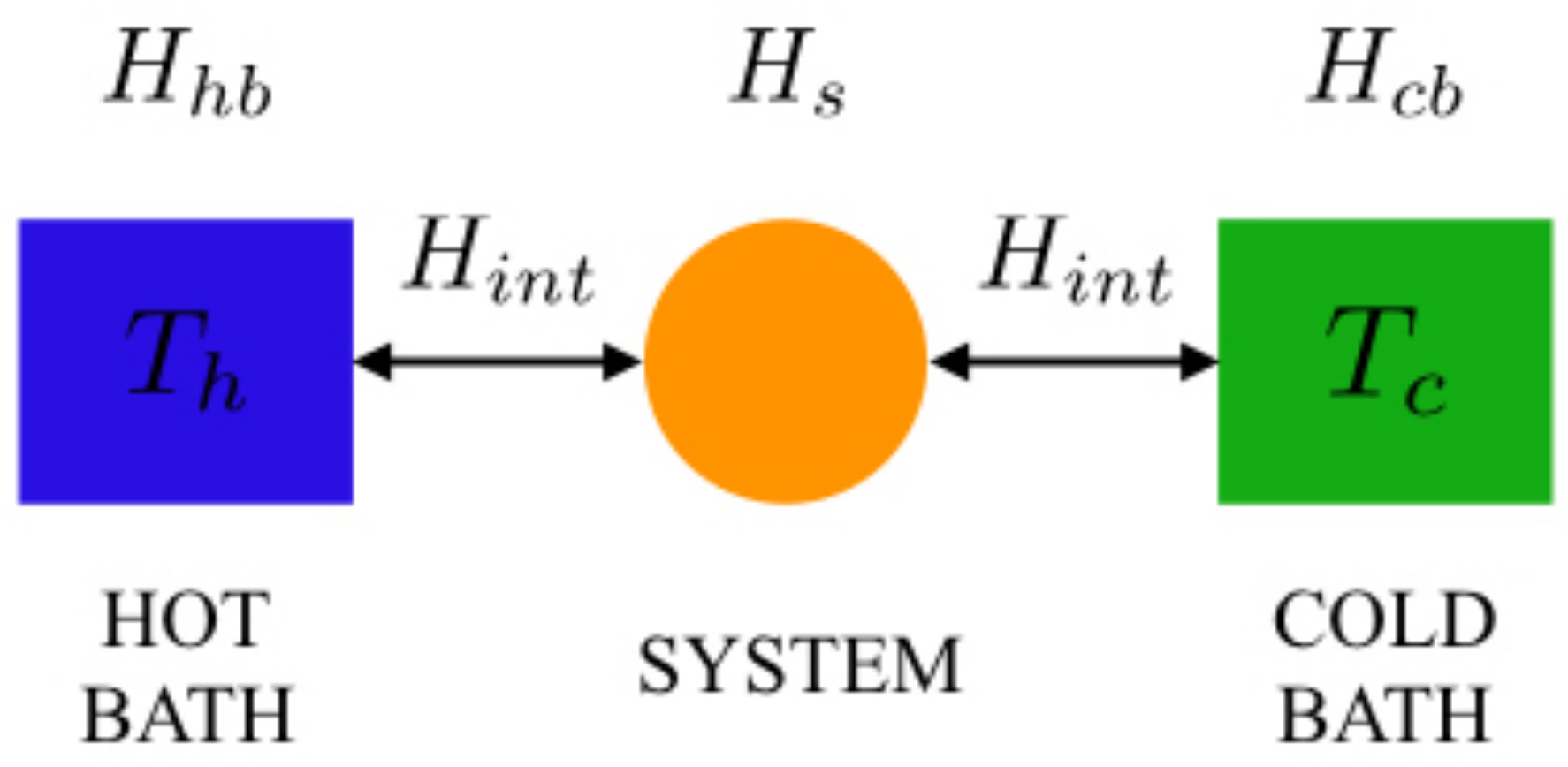}
\caption{(Color online) Composite system.}
\label{cartoon}
\end{center}
\end{figure}
The total Hamiltonian is given by
 \begin{align}
 H_{tot}(z,y_c,y_h;\lambda(t))&= H_s(z,\lambda(t))+H_{hb}(y_h)+H_{cb}(y_c)+H_{int}(z,y_h,y_c).
 \end{align}
Here, $H_s$ is the Hamiltonian of the system, $H_{hb}$ and $H_{cb}$ are Hamiltonians for the hot and the cold baths, respectively.  $H_{int}$ consists of the combined interaction Hamiltonians with the hot and the cold baths. $\lambda(t)$ is an external parameter that can be changed  according to a fixed protocol. In our analysis, we will assume that the interaction energy is negligible compared to the energies of the heat baths and of the system. The interaction Hamiltonian is introduced only to ensure that the total system follows a Hamiltonian dynamics. The system is assumed to begin and end in noneqilibrium steady states.

 We now define the thermodynamic quantities. The work done on the system must be equal to the change in internal energy of the full system (consisting of the baths and the system), because the total energy can change only because of work being done on the system of interest. Thus, we can write
 \begin{align}
 W&= H_{tot}(\tau)-H_{tot}(0)=\int_0^\tau \partial_t H_s(t) dt,
 \end{align}
 where $\tau$ is the time of observation, and the arguments of $H_{tot}$ and $H_s$ have been suppressed for notational simplicity. The last equality comes from the fact that only the Hamiltonian of the system is explicitly dependent on time.  The heat dissipated by the system into any of the heat baths, on the other hand, must be equal to the change in the energy of the corresponding bath, so that we have
 \begin{align}
 Q_h = H_{hb}(\tau)-H_{hb}(0)\hspace{1 cm} Q_c = H_{cb}(\tau)-H_{cb}(0).
 \end{align}
 The change in energy of the system is given by the change in the system's Hamiltonian:
 \begin{align}
 \Delta E = H_s(\tau)-H_s(0).
 \end{align}
The First law of thermodynamics is 
\begin{equation}
W=Q_h+Q_c+\Delta E.
\label{1st law}
\end{equation}
The hot and cold reservoirs are initially prepared at temperatures $T_h$ and $T_c$ respectively, i.e., initial microstates $y_h(0)$ and $y_c(0)$ are sampled from canonical ensembles:
\begin{align}
\label{dist}
 P_h(y_{h}(0)) &= \frac{e^{-\beta_h H_{hb}(y_{h}(0))}}{Z_h}; \hspace{1cm} P_c(y_{c}(0)) = \frac{e^{-\beta_c H_{cb}(y_{c}(0))}}{Z_c}.
 \end{align}
Here, $Z_h$ and $Z_c$ are the canonical partition functions for the hot and the cold baths, respectively. The system is initially prepared at equilibrium with the cold bath, so that its initial distribution is
\begin{align}
  p_s(z(0))&= \frac{e^{-\beta_c H_s(z(0))}}{Z_s(0)}.
\end{align}
At time $t=0$, we assume that  the probability distribution of combined system (system+heat baths) are in product form. At time $t=0^+$,  the external protocol $\lambda(t)$ is switched on and it continues for a time $\tau'$, after which the hot bath is disconnected and the system is allowed to equilibrate with the cold bath. Let the final time (after equilibration) be $\tau$. During this entire time the composite system undergoes a Hamiltonian evolution (forward process denoted by $\Pi^+$) from the initial state $\Gamma=(z(0),y_h(0),y_c(0))$ and let 
\begin{align}
\hat\Gamma^t_+(\Gamma)\equiv(\hat z^t_+(\Gamma), \hat y^t_{h+}(\Gamma), \hat y^t_{c+}(\Gamma))
\end{align}
denote the point in phase space reached after time $t$. At $t=\tau$, the final microstate of the composite system in the full phase space is $\Gamma'=\hat\Gamma^{\tau}_+(\Gamma)\equiv(\hat z^{\tau}_+(\Gamma), \hat y^{\tau}_{h+}(\Gamma), \hat y^{\tau}_{c+}(\Gamma))=(z',y_h',y_c')$. 

%Let $p_s(z')$ denote the final distribution of the system. The initial and the final entropies of the system in the forward process are given by
%
 %\begin{align}
 %S(0)=-\ln p_s(z_0); \hspace{1cm} S(\tau)=-\ln p_s(z'),
 %\end{align}
 %
%and the change in entropy during the process is given by 
%
%\begin{align}
  %\Delta S \equiv S(\tau)-S(0) = \ln\frac{p_s(z_0)}{p_s(z')}.
%\end{align}

\subsection{Derivation of the central result}
\label{sec:central}
We now calculate the joint probability $P_+(Q_h,\Delta E,W)$ in the forward process.  Let $\hat Q_h^+(\Gamma), \Delta \hat E^+(\Gamma)$ and $\hat W^+(\Gamma)$ denote respectively the heat generated, internal energy change and work done on the system over the entire realisation of the process (from $t=0$ to $t=\tau$). For the time reversed process (denoted by $\Pi^-$), we adopt the same notation with a change in the subscript ($\hat\Gamma^t_-,\hat z_-,$ etc.). Then we have,
 \begin{eqnarray}
 P_+(Q_h,\Delta E,W)&=&\langle \delta(Q_h-\hat Q_h^+(\Gamma))\delta(\Delta E-\Delta \hat E^+(\Gamma))\delta(W-\hat W^+(\Gamma))\rangle\nonumber\\
 &=&\int dz(0)dy_h(0)dy_c(0) p_s(z(0))P_h(y_{h}(0))P_c(y_{c}(0))\nonumber\\
 &&\hspace{.5cm} \times \delta(Q_h-\hat Q_h^+(\Gamma))\delta(\Delta E-\Delta \hat E^+(\Gamma))\delta(W-\hat W^+(\Gamma)).
 \label{joint}
 \end{eqnarray} 
Let us define two normalized functions $P_h(y_h')$ and $P_c(y_c')$, which have the canonical form as in Eq. \eqref{dist}, evaluated at $y_h'=\hat y^{\tau}_{h+}(\Gamma)$ and $y_c'=\hat y^{\tau}_{c+}(\Gamma)$ respectively. %Note that these function are \emph{not} the true distributions of the bath states at the end of the process.
Now we can rewrite $p_s(z(0)), P_h(y_{h}(0))$ and $P_c(y_{c}(0))$ as 
\begin{eqnarray}
p_s(z(0))&=&e^{\beta_c [H_s(z')-H_s(z(0))]}\frac{Z_s(\tau)}{Z_s(0)}\frac{e^{-\beta_c H_s(z')}}{Z_s(\tau)}=e^{\beta_c(\Delta E-\Delta F)}\frac{e^{-\beta_c H_s(z')}}{Z_s(\tau)},\\
P_h(y_{h}(0))&=&\frac{P_h(y_{h}(0))}{P_h(y_{h}')}P_h(y_{h}')=e^{\beta_hQ_h}P_h(y_{h}'), \\
P_c(y_{c}(0))&=&\frac{P_c(y_{c}(0))}{P_c(y_{c}')}P_c(y_{c}')=e^{\beta_cQ_c}P_c(y_{c}'), 
\end{eqnarray}
where $\beta_h$ and $\beta_c$ are the inverse temperature of the hot and the cold bath respectively.  Substituting these expressions in Eq. \eqref{joint} and using the fact that the Jacobian between $(z(0),y_h(0),y_c(0))$ and $(z',y_h',y_c')$ is unity (by Liouville's theorem), we have
\begin{eqnarray}
 P_+(Q_h,\Delta E,W)&=&\int dz'dy_h'dy_c' \exp[\beta_hQ_h+\beta_c(Q_c+\Delta E-\Delta F)]\frac{e^{-\beta_c H(z')}}{Z_s(\tau)}P_h(y_{h}')P_c(y_{c}')\nonumber\\
 &&\hspace{2cm} \times \delta(Q_h-\hat Q_h^+(\Gamma))\delta(\Delta E-\Delta \hat E^+(\Gamma))\delta(W-\hat W^+(\Gamma)).
 \label{joint1}
\end{eqnarray}
Now if we reverse the final momenta and start a realisation of $\Pi^-$ from the initial conditions $\Gamma'^*$ , we will obtain the time-reversed image of the original realisation: $\hat\Gamma^t_-(\Gamma'^*)=[\hat\Gamma^{\tau-t}_+(\Gamma)]^*$. In the reverse process initially the bath microstates are chosen from the canonical ensemble given by
\begin{align}
\label{dist1}
 P_h(y_{h}'^*) &= \frac{e^{-\beta_h H_{hb}(y_{h}'^*)}}{Z_h}=\frac{e^{-\beta_h H_{hb}(y_{h}')}}{Z_h}=P_h(y_{h}');~~ P_c(y_{c}'^*) = \frac{e^{-\beta_c H_{cb}(y_{c}'^*)}}{Z_c}=\frac{e^{-\beta_c H_{cb}(y_{c}')}}{Z_c}=P_c(y_{c}'),
 \end{align}
 and the initial system microstate is chosen from 
 \begin{align}
   p_s(z'^*)=\frac{e^{-\beta_c H(z'^*)}}{Z_s(\tau)}=\frac{e^{-\beta_c H(z')}}{Z_s(\tau)}.
 \end{align}
The second equalities in each of the above two equations can be written due to invariance of the Hamiltonians under time-reversal. From time reversal it follows that 
 \begin{align}
 \hat Q_h^-(\Gamma'^*)=-\hat Q_h^+(\Gamma); \hspace{.3cm}\Delta \hat E^-(\Gamma'^*)=-\Delta \hat E^+(\Gamma);\hspace{.3cm}\hat W^-(\Gamma'^*)=-\hat W^+(\Gamma).
 \end{align}
 %
%Note that the assumption of a steady state at the beginning as well at the end of the process is necessary to ensure that the system entropy change switches sign under time-reversal \cite{cro99_pre}.
 The above relations allow us to rewrite Eq. \eqref{joint1} as 
 \begin{eqnarray}
 P_+(Q_h,\Delta E,W)&=&\int dz'^*dy_h'^*dy_c'^* \exp[\beta_hQ_h+\beta_c(Q_c+\Delta E-\Delta F)]p_s(z'^*)P_h(y_{h}'^*)P_c(y_{c}'^*)\nonumber\\
 &&\hspace{2cm} \times \delta(Q_h+\hat Q_h^-(\Gamma'^*))\delta(\Delta E+\Delta \hat E^-(\Gamma'^*))\delta(W+\hat W^-(\Gamma'^*)).
 %&=&e^{\beta_hQ_h+\beta_cQ_c+\Delta S}P_-(-Q_h,-\Delta E,-W,-\Delta S)
 \end{eqnarray}
 The change of variable from $(z',y_h',y_c')$ to $(z'^*,y_h'^*,y_c'^*)$ can be done due to one-to-one correspondence between $\Gamma'$ and $\Gamma'^*$. We now use the first law Eq. \ref{1st law} to rewrite the above equation as
 \begin{align}
   P_+(Q_h,\Delta E,W)&=\exp[(\beta_h-\beta_c)Q_h+\beta_c(W-\Delta F)]P_-(-Q_h,-\Delta E,-W).
   \label{joint_qh}
\end{align}
Integrating over all values of $\Delta E$ we obtain
\begin{align}
\frac{P_+(Q_h,W)}{P_-(-Q_h,-W)} &= \exp\big[(\beta_h-\beta_c)Q_h+\beta_c(W-\Delta F)\big],
 \label{eq:DFT}
\end{align}
which is our first main result. This is the generalised detailed FT. One now immediately gets
\begin{align}
  \frac{P_+(Q_h)}{P_-(-Q_h)}=\frac{e^{(\beta_h-\beta_c)Q_h}}{\Psi(Q_h)},
  \label{eq:DFT_modified}
\end{align}
where
\begin{align}
  \Psi(Q_h) = \langle e^{-\beta_c(W-\Delta F)}\big|Q_h\rangle.
\end{align}
The average on the right hand side means that the averaging has been done over the conditional probability distribution $P_+(W|Q_h)$. %In other words, it is an average over all the forward trajectories having a given value of $Q_h$.
Thus, we obtain the integral FT
\begin{align}
  \langle e^{-(\beta_h-\beta_c)Q_h+\ln\Psi(Q_h)}\rangle=1,
\end{align}
from which we obtain using Jensen's inequality \cite{jensen}
\begin{align}
  (\beta_h-\beta_c)\langle Q_h\rangle\ge \langle\ln\Psi(Q_h)\rangle.
\end{align}
We observe that if  we have $\Psi(Q_h)\ge 1$ then $\av{Q_h}$ must be negative, so that the system absorbs heat on average from the hot bath. This differs from the earlier work by Jarzynski and W$\grave{o}$jcik by a factor of $\Psi(Q_h)$ in the presence of external drive. If the external drive is switched off, we recover the J-W result in the transient state.In the absence of external drive $W=0$ and consequently $\Psi(Q_h)=1$. We get the result
\begin{equation}
\la e^{-(\beta_h-\beta_c)Q_h}\ra=1.
\end{equation}
Using Jensen's inequality we have $(\beta_h-\beta_c)\la Q_h\ra\geq 0$, which is consistent with the second law of thermodynamics. It is important to state that our results are valid only transient state and not for the steady states as done in \cite{lah14_epjb}.
%Similarly, one can show that if $\Psi(Q_c)\ge 1$, then the system dissipates heat on average into the cold bath.

%One can also find a heat fluctuation theorem for $Q_c$. In this case we start with the joint probability distribution $P(Q_c,\Delta E,W)$.  Following the same steps as before we obtain the following equation:
%
%\begin{eqnarray}
 %P_+(Q_c,\Delta E,W)&=&\int dz'^*dy_h'^*dy_c'^* \exp[\beta_hQ_h+\beta_c(Q_c+\Delta E-\Delta F)]p_s(z'^*)P_h(y_{h}'^*)P_c(y_{c}'^*)\nonumber\\
 %&&\hspace{2cm} \times \delta(Q_c+\hat Q_c^-(\Gamma'^*))\delta(\Delta E+\Delta \hat E^-(\Gamma'^*))\delta(W+\hat W^-(\Gamma'^*)).
 %\end{eqnarray}
%Using First law ($Q_h=W-Q_c-\Delta E$), the above equation can be rewritten as 

%\begin{equation}
%P_+(Q_c,\Delta E,W)=\exp[(\beta_c-\beta_h)(Q_c+\Delta E)+\beta_h W-\beta_c \Delta F]P_-(-Q_c,-\Delta E,-W).
%\label{joint_qc}
%\end{equation}
 
%Corresponding heat fluctuation theorem is given by
%\begin{align}
%\frac{P_+(Q_c)}{P_-(-Q_c)}=\frac{e^{(\beta_c-\beta_h)Q_c}}{\tilde{\Psi}(Q_c)},
%\end{align}
%where $\tilde{\Psi}(Q_c)=\int \exp[-(\beta_c-\beta_h)\Delta E-\beta_h W+\beta_c \Delta F)]P_+(\Delta E,W|Q_c)dWd(\Delta E)$.  
 
 \section{Generalization to the quantum case}
\label{sec:quantum}
\subsection{System connected to single heat bath}
The above case can be generalised to quantum systems. To define the thermodynamic quantities in this case, we first note that work is not a quantum observable, so we need to perform projective measurements on the composite system (system+heat baths) at the beginning and at the end of the protocol, as described in \cite{han11_rmp}. Initially, the system is prepared in equilibrium with the heat bath at temperature $\beta^{-1}$.  At time $t = 0$, we perform a projective measurement on the Hamiltonians of the system and bath. Since the bath and the system hamiltonian commute, their eigenstates can be simultaneously measured. Let $|i_0\rangle\equiv|n,\nu\rangle$ denote the combined state of the system and the bath. The corresponding energy eigenvalues of the system and the bath are $(E_n,E_{\nu})$. The total system (system of interest+baths) then evolve unitarily under the action of an external drive $\lambda(t)$ up to time $t=\tau$. At the final time, we once again perform projective measurements simultaneously on the system and the bath.  Let $|i_{\tau}\rangle\equiv|m,\mu\rangle$ denote the combined state of the system and the bath at the end of the process. The change in energy of the system is given by
\begin{align}
  \Delta E&= E_m-E_n.
\end{align}
The heat dissipated into the bath in the process is 
\begin{align}
  Q &= E_{\mu}-E_{\nu}.
  \label{q_q}
\end{align}
The work done on the system is obtained from the First law as 
\begin{eqnarray}
W=\Delta E+Q=E_m-E_n+E_{\mu}-E_{\nu}.
\end{eqnarray}
The initial density operator for the full system is
\begin{align}
  \hat\rho(0)= \frac{e^{-\beta \hat H_s(0)}}{Z_s(0)}\otimes \frac{e^{-\beta \hat H_b(0)}}{Z_b}.
\end{align}
The probability of obtaining the eigenstate $|i_0\rangle$ is then given by
\begin{align}
  P(i_0)\equiv P(n,\nu)=p_s(n)P_b(\nu)= \frac{e^{-\beta E_n}}{Z_s(0)}\cdot\frac{e^{-\beta E_{\nu}}}{Z_b}.
\end{align}
We now calculate the forward joint probability distribution $P_+(Q,\Delta E)$. Let  $Q^+(\mu,\nu)$ and $ \Delta E^+(m,n)$ denote the heat flows to the hot bath and change in internal energy of the system in the forward process, respectively. For the reverse process we adopt the same notation with the $'+'$ sign in the subscript replaced by $'-'$ and the arguments are marked with a $*$, which implies action of time-reversal operator on the states. Then we have

\begin{eqnarray}
P_+(Q,\Delta E)&=& \sum_{i_0,i_{\tau}}P(i_0,i_{\tau})\delta(Q-Q^+(\mu,\nu))\delta(\Delta E-\Delta E^+(m,n))\nonumber\\
&=& \sum_{i_0,i_{\tau}}P(i_{\tau}|i_0)P(i_0)\delta(Q-Q^+(\mu,\nu))\delta(\Delta E-\Delta E^+(m,n)).
\label{joints_q}
\end{eqnarray}

Here $P(i_{\tau}|i_0)$ is the transition probability given by
\begin{align}
P(i_{\tau}|i_0)=|\langle i_{\tau}|\hat U_{\lambda(t)}(\tau,0)|i_0\rangle|^2,
\end{align}
where $\hat U_{\lambda(t)}(\tau,0)$ is the unitary operator that changes the state of the combined system from $|i_0\rangle$ to $|i_{\tau}\rangle$ in time $\tau$ under the action of the protocol $\lambda(t)$.

In the time reversed process the combined system starts from a state given by  

\begin{align}
 \hat \rho^*(\tau)= \frac{e^{-\beta \hat H^*_s(\tau)}}{Z_s(\tau)}\otimes \frac{e^{-\beta \hat H^*_b(\tau)}}{Z_b}=\frac{e^{-\beta \hat H_s(\tau)}}{Z_s(\tau)}\otimes \frac{e^{-\beta \hat H_b(\tau)}}{Z_b},
\end{align}
where we have used the time reversal invariance of the Hamiltonian operators in the last equation.  The probability of obtaining the eigenstate $|i^*_{\tau}\rangle\equiv|m^*,\mu^*\rangle$ is 
\begin{eqnarray}
P(i_{\tau}^*)\equiv P(m^*)P(\mu^*)&=&\frac{e^{-\beta E_m}}{Z_s(\tau)}\frac{e^{-\beta E_{\mu}}}{Z_b}=e^{-\beta (E_m-E_n)}\frac{Z_s(0)}{Z_s(\tau)}\frac{e^{-\beta E_n}}{Z_s(0)}e^{-\beta (E_{\mu}-E_{\nu})}\frac{e^{-\beta E_{\nu}}}{Z_b}\nonumber\\
&=&e^{-\beta (Q+\Delta E-\Delta F)}p_s(n)P_b(\nu)=e^{-\beta (Q+\Delta E-\Delta F)}P(i_0).
\label{ris_q}
\end{eqnarray}
In the reverse process the system ends in the state $|i^*_{0}\rangle=|n^*,\nu^*\rangle$. The transition probability from $|i^*_{\tau}\rangle$ to $|i^*_{0}\rangle$ is 
\begin{align}
P(i^*_0|i^*_{\tau})=|\langle i^*_0|\hat U^{\dag}_{\lambda^*(t)}(\tau,0)|i^*_{\tau}\rangle|^2=|\langle i_{\tau}|\hat U_{\lambda(t)}(\tau,0)|i_0\rangle|^2=P(i_{\tau}|i_0),
\label{tr_q}
\end{align}
where $\lambda^*(t)=\lambda(\tau-t)$ is the time reversed protocol. Under the application of time reversed protocol, all the thermodynamic quantities acquire a negative sign, i.e.,
 \begin{align}
 Q^-(\mu^*,\nu^*)=-Q^+(\mu,\nu);\hspace{.3cm}\Delta E^-(m^*,n^*)=-\Delta E^+(m,n).
 \label{qs_q}
 \end{align}
Using Eqs. \ref{ris_q}, \ref{tr_q} and \ref{qs_q}, Eq. \ref{joints_q} can be rewritten as 
\begin{eqnarray}
P_+(Q,\Delta E)&=& \sum_{i^*_0,i^*_{\tau}}P(i^*_0|i^*_{\tau})e^{\beta (Q+\Delta E-\Delta F)}P(i_{\tau}^*)\delta(Q+Q^-(\mu^*,\nu^*))\delta(\Delta E+\Delta E^-(m^*,n^*))\nonumber\\
&=&e^{\beta (Q+\Delta E-\Delta F)}P_-(-Q,-\Delta E).
\end{eqnarray}
Summation over all possible values of $\Delta E$ leads to the following quantum heat FT
\begin{equation}
\frac{P_+(Q)}{P_-(-Q)}=\frac{e^{\beta Q}}{\Psi(Q)},
\end{equation}
where 
\begin{align}
\Psi(Q)=e^{\beta \Delta F}\sum_{\Delta E}e^{-\beta \Delta E}P_+(\Delta E|Q).
\end{align}
This has the same form obtained classically (Eq.\ref{psi_classical}).

\subsection{System connected to two heat baths}
We assume that initially the two baths have been equilibrated, so that their density matrices are the canonical ones. At time $t = 0$, we perform a projective measurement on the Hamiltonians of the system and the two baths. Since the system and the bath Hamiltonians commute with each other, their eigenstates can be simultaneously measured. Let $|i_0\rangle\equiv|n,\nu_h,\nu_c\rangle$ denote the combined state of the system and the two baths (subscripts $h$ and $c$ represent hot and cold bath respectively). The corresponding energy eigenvalues of the system and the baths be $(E_n,E_{\nu_h},E_{\nu_c})$. The total system  then evolve unitarily under the action of an external drive $\lambda(t)$ up to time $t=\tau$. At the final time, we once again perform projective measurements simultaneously on the system and the baths.  Let $|i_{\tau}\rangle\equiv|m,\mu_h,\mu_c\rangle$ denote the combined state of the system and the two baths at the end of the process. The change in energy of the system is given by
\begin{align}
  \Delta E&= E_m-E_n.
  \label{ed_q}
\end{align}
The heat dissipated into the hot bath during the process is 
\begin{align}
  Q_h &= E_{\mu_h}-E_{\nu_h},
  \label{qh_q}
\end{align}
and that into the cold bath is
\begin{align}
  Q_c &= E_{\mu_c}-E_{\nu_c}.
  \label{qc_q}
\end{align}
The work done on the system is obtained from the First law as 
\begin{eqnarray}
W=\Delta E+Q_h+Q_c=E_m-E_n+E_{\mu_h}-E_{\nu_h}+E_{\mu_c}-E_{\nu_c}.
\end{eqnarray}
We note that the work done is equal to the change in total energy (neglecting the interaction terms) of the combined system during the process.
%The change in system's entropy during the process is 
%
%\begin{align}
  %\Delta S= \ln\frac{p_s(n)}{p_s(m)},
  %\label{ds_q}
%\end{align}
%
%where $p_s(m)$ is the probability of the final state of the system. 
As before, we assume the system to be at equilibrium with the cold bath at the beginning of the process. The initial density operator for the full system is
\begin{align}
  \hat \rho(0)= \frac{e^{-\beta_c \hat H_s(0)}}{Z_s(0)}\otimes \frac{e^{-\beta_h \hat H_h(0)}}{Z_h}\otimes \frac{e^{-\beta_c\hat H_c(0)}}{Z_c}.
\end{align}
The probability of obtaining the eigenstate $|i_0\rangle$ is then given by
\begin{align}
  P(i_0)\equiv P(n,\nu_h,\nu_c)=p_s(n)P_h(\nu_h)P_c(\nu_c)= \frac{e^{-\beta_c E_n}}{Z_s(0)}\cdot\frac{e^{-\beta_h E_{\nu_h}}}{Z_h}\cdot\frac{e^{-\beta_c E_{\nu_c}}}{Z_c}.
\end{align}
We now calculate the forward joint probability distribution $P_+(Q_h,\Delta E, W)$. Let  $Q_h^+(\mu_h,\nu_h), \Delta E^+(m,n)$ and $W^+(i_0,i_{\tau})$  denote the heat flow to the hot bath, change in internal energy of the system and work done on the system in the forward process respectively. For the reverse process we adopt the same notation with the $'+'$ sign in the subscript replaced by $'-'$ and the arguments are marked with a $*$. Then we have
\begin{eqnarray}
  P_+(Q_h,\Delta E, W) &=& \sum_{i_0,i_{\tau}}P(i_0,i_{\tau})\delta(Q_h-Q_h^+(\mu_h,\nu_h))\delta(\Delta E-\Delta E^+(m,n))\delta(W-W^+(i_0,i_{\tau}))\nonumber\\
  &=&\sum_{i_0,i_{\tau}}P(i_{\tau}|i_0)P(i_0)\delta(Q_h-Q_h^+(\mu_h,\nu_h))\delta(\Delta E-\Delta E^+(m,n))\delta(W-W^+(i_0,i_{\tau})).
 \label{joint_q}
\end{eqnarray}
Here $P(i_{\tau}|i_0)$ is the transition probability given by
\begin{align}
P(i_{\tau}|i_0)=|\langle i_{\tau}|\hat U_{\lambda(t)}(\tau,0)|i_0\rangle|^2,
\end{align}
where $\hat U_{\lambda(t)}(\tau,0)$ is the unitary operator that changes the state of the combined system from $|i_0\rangle$ to $|i_{\tau}\rangle$ in time $\tau$ under the action of the protocol $\lambda(t)$.

In the time reversed process the combined system starts from a state given by  
\begin{align}
 \hat  \rho^*(\tau)= \frac{e^{-\beta_c \hat H_s(\tau)}}{Z_s(\tau)}\otimes \frac{e^{-\beta_h \hat H^*_h(\tau)}}{Z_h}\otimes \frac{e^{-\beta_c\hat H^*_c(\tau)}}{Z_c}.
\end{align}
Since the forward process ends with the system being in the state $|i_{\tau}\rangle$, the reverse process will start from the time reversed state of this state i.e., $|i^*_{\tau}\rangle=\hat \Theta|i_{\tau}\rangle$, where $\hat \Theta $ is the time-reversal operator. The probability of obtaining the eigenstate $|i^*_{\tau}\rangle\equiv|m^*,\mu^*_h,\mu^*_c\rangle$ is 
\begin{eqnarray}
P(i^*_{\tau})\equiv P(m^*,\mu^*_h,\mu^*_c)&=&p_s(m^*)P_h(\mu^*_h)P_c(\mu^*_c)\nonumber\\
&=&\exp[-\beta_h Q_h-\beta_c Q_c-\beta_c(\Delta E-\Delta F)]p_s(n)P_h(\nu_h)P_c(\nu_c)\nonumber\\
&=&\exp[-\beta_h Q_h-\beta_c Q_c-\beta_c(\Delta E-\Delta F)]P(i_0).
\label{ri_q}
\end{eqnarray}

The above equation is obtained by using the definitions given in Eqs. \eqref{ed_q}, \eqref{qh_q} and \eqref{qc_q}  and the fact that density operator of the heat baths and system are time reversal invariant at time $t=\tau$.
In the reverse process the system ends in the state $|i^*_{0}\rangle=\Theta|i_{0}\rangle$. The transition probability from $|i^*_{\tau}\rangle$ to $|i^*_{0}\rangle$ is 
\begin{align}
P(i^*_0|i^*_{\tau})=|\langle i^*_0|\hat U^{\dag}_{\lambda^*(t)}(\tau,0)|i^*_{\tau}\rangle|^2=|\langle i_{\tau}|\hat U_{\lambda(t)}(\tau,0)|i_0\rangle|^2=P(i_{\tau}|i_0),
\label{rt_q}
\end{align}
where $\lambda^*(t)=\lambda(\tau-t)$ is the time reversed protocol. Under the application of time reversed protocol, all the thermodynamic quantities acquire a negative sign, i.e.,
 \begin{eqnarray}
&& Q_h^-(\mu_h^*,\nu_h^*)=-Q_h^+(\mu_h,\nu_h); \hspace{.3cm}\Delta E^-(m^*,n^*)=-\Delta E^+(m,n);\nonumber\\
&&\hspace{3cm} W^-(i_0^*,i_{\tau}^*)=- W^+(i_0,i_{\tau}).
 \label{quantities}
 \end{eqnarray}
Substituting Eqs. \eqref{ri_q}, \eqref{rt_q} and \eqref{quantities} in Eq. \eqref{joint_q} we obtain
\begin{eqnarray}
P_+(Q_h,\Delta E, W) &=& \sum_{i_0^*,i^*_{\tau}}P(i_0^*|i_{\tau}^*)\exp[\beta_h Q_h+\beta_c Q_c+\beta_c(\Delta E-\Delta F)]P(i_{\tau}^*)\delta(Q_h+Q_h^-(\mu^*_h,\nu^*_h))\nonumber\\
  &&\hspace{2 cm}\times \delta(\Delta E+\Delta E^-(m^*,n^*))\delta(W+W^-(i^*_0,i^*_{\tau}))
\end{eqnarray}

Replacing $Q_c$ from the First law ($W=\Delta E+Q_h+Q_c$) we arrive at the following:
\begin{eqnarray}
P_+(Q_h,\Delta E, W) 
  &=&\exp[(\beta_h-\beta_c) Q_h+\beta_c(W-\Delta F)]P_-(-Q_h,-\Delta E, -W).
\end{eqnarray}
Summation on both sides of the above equation over all possible values of $\Delta E$ and $W$ leads to the following fluctuation relation for $Q_h$ in the transient regime:
\begin{align}
 \frac{P_+(Q_h)}{P_-(-Q_h)}=\frac{e^{(\beta_h-\beta_c)Q_h}}{\Psi(Q_h)},
\end{align}
where 
\begin{align}
\Psi(Q_h)=\sum_{W}e^{-\beta_c(W- \Delta F)}P_+(W|Q).
\end{align}
In absence of the protocol $W=0$ and consequently $\Psi(Q_h)=1$. Hence, we get 
\begin{equation}
\frac{P_+(Q_h)}{P_-(Q_h)}=e^{(\beta_h-\beta_c)Q_h}.
\end{equation}
The corresponding integral FT is 
\begin{equation}
\la e^{-(\beta_h-\beta_c)Q_h}\ra=1.
\end{equation}
Using Jensen's inequality we have $(\beta_h-\beta_c)\la Q_h\ra\geq 0$, which is consistent with the second law of thermodynamics.
% 
%

% \section{Case of heat engines}

% The above analysis holds for microscopic heat engines both in the classical and in the quantum regimes. Note that the derivation in the quantum case has already been provided in \cite{cam14_jpa}. The explicit derivation for classicial case in Hamiltonian framework is missing, although heat engines governed by stochastic jump processes has been worked out in \cite{sin11_jpa}. 

% When the full system including the heat engine and the two baths is described by Hamiltonian dynamics, we can follow the steps in sec. \ref{sec:central} and arrive at Eq. \eqref{eq:DFT}. The corresponding integral FT can be readily derived:
% %
% \begin{align}
%   \av{\exp[-(\beta_h-\beta_c)Q_h-\beta_c(W-\Delta E)-\Delta S]}=1.
% \end{align}
% %
% We can now apply the Jensen's inequality to the above equation and obtain
% %
% \begin{align}
%   (\beta_h-\beta_c)\av{Q_h}+\beta_c(\av{W}-\av{\Delta E})+\av{\Delta S}\ge 0.
% \end{align}
% %
% Since the engine undergoes a cyclic process, we can 

\section{Conclusions}
\label{sec:conclusion}

In this work, our focus has been on a system that is simultaneously connected to two heat baths at different temperatures. At first, we use the treatment to describe a system connected to a single bath. The composite system consisting of the heat bath and the system of interest follow Hamiltonian dynamics, and the interaction term between the system and the bath is considered to be negligible. The joint probabilities of the dissipated heat ($Q$) and the change in internal energy ($\Delta E$) are shown to follow a detailed fluctuation relation. We have then extended the treatment to the case where the system is simultaneously connected to two heat baths. In this case, the joint probabilities of the heat dissipated into the hot bath ($Q_h$), the work done on the system ($W$) and change in internal energy of the system ($\Delta E$), in the forward and reverse processes, are related through the fluctuation theorem, Eq. \eqref{joint_qh}. This leads to a modified FT for $Q_h$, where a factor $\Psi(Q_h)$ appears (see Eq. \eqref{eq:DFT_modified}) that quantifies the deviation of the symmetry function from a straight line. This is consistent with \cite{noh12_prl}. Same results are obtained in quantum regime. For the system connected to two heat baths we obtained modified J-W FT in both classical and quantum cases. In this work we have assumed weak coupling between the baths and the intermediate medium. Presently, we are extending our results to systems where coupling is strong \cite{talkner16_ar} and separately performing weak measurement on the composite system.

\section{Acknowldgement}
A.M.J thanks Department of Science and Technology, India, for financial support (through J. C. Bose National Fellowship).

%\section*{References}

\end{document}